# Electric-Field Induced Superconductor-Insulator Transitions in Exfoliated Bi2Sr2CaCu2O8+δ Flakes

Menghan Liao[1], Yuying Zhu[1], Jin Zhang[1], Ruidan Zhong[2], John Schneeloch[2,3], Gen-Da Gu[2], Kaili Jiang[1,4,5], Ding Zhang[1,5 #], Xu-Cun Ma[1,5] and Qi-Kun Xue[1,5 #]

[1]State Key Laboratory of Low Dimensional Quantum Physics and Department of Physics, Tsinghua University, Beijing, 100084, China

[2]Condensed Matter Physics and Materials Science Department, Brookhaven National Laboratory, Upton, New York 11973, USA

[3]Department of Physics and Astronomy, Stony Brook University, Stony Brook, New York 11794, USA

[4]Tsinghua-Foxconn Nanotechnology Research Center, Tsinghua University, Beijing, 100084, China

[5]Collaborative Innovation Center of Quantum Matter, Beijing, 100084, China

**ABSTRACT:** We realize superconductor-insulator transitions (SIT) in mechanically exfoliated $Bi_2Sr_2CaCu_2O_{8+\delta}$ (BSCCO) flakes and address simultaneously their transport properties as well as the evolution of density of states. Back-gating via the solid ion conductor engenders a reversible SIT in BSCCO, as lithium ions from the substrate are electrically driven into and out of BSCCO. Scaling analysis indicates that the SIT follows the theoretical description of a two-dimensional quantum phase transition (2D-QPT). We further carry out tunneling spectroscopy in graphite(G)/BSCCO heterojunctions. We observe V-shaped gaps in the critical regime of the SIT. The density of states in BSCCO gets symmetrically suppressed by further going into the insulating regime. Our technique of combining solid state gating with tunneling spectroscopy can be easily applied to the study of other two-dimensional materials.

Cuprate superconductors exhibit a rich variety of correlated phases that can be traversed by tuning a single parameter–the charge carrier density[1, 2]. Understanding the emergence of these phases and their competition/coexistence with superconductivity is of vital importance, as it may shed light on the mechanism for the high transition temperature of the cuprate superconductors. Of late, ionic liquid gating has been proven especially effective in modulating the charge carrier density[3, 4], switching the cuprates between a superconductor and an insulator[5–9]. SIT has been realized in both hole-doped ($La_{2-x}Sr_xCuO_4$[5], $YBaCu_3O_{7-x}$[6], $La_2CuO_{4+\delta}$[8]) and electron-doped ($Pr_{2-x}Ce_xCuO_4$[9]) cuprates with the superconducting block of only 1 to 4 unit cells. These studies reveal that the SIT in cuprates are 2D-QPT. Further experiments demonstrated a continuous tuning from the underdoped regime all the way into the overdoped regime, revealing possible electronic phase transition at the optimal doping[7].

This electric-field induced continuous tuning of the doping level in cuprates could be highly beneficial for investigating the complex phase diagram[10, 11]. The fine tuning across the quantum critical region, for example, can help gaining further insight into the competition between charge density waves and superconductivity[12]. In contrast, present state-of-the art spectroscopic studies need to address samples with different doping levels sequentially[10, 13, 14]. However, the evolution of electronic states under liquid gating remains challenging to be explored. The sample surface gets masked by the liquid, which is incompatible with ultrahigh high vacuum. The liquid may also flow away once the sample stage is tilted. Recently, the solid ion conductor (SIC) emerges as an effective alternative for charge carrier modulation[15]. The first realization of a SIC field effect transistor, which was achieved on FeSe flakes, demonstrated dramatic enhancement of superconductivity[15]. The SIC was employed not only for electrostatic tuning but also for electrically driven intercalation of lithium ions. Following-up experiments were carried out on graphene and showed comparable electrostatic tunability to that of the ionic liquid/gel[16, 17]. Furthermore, the solid state gating is compatible with spectroscopic investigations[15] such that complex electronic phases in the phase diagram of cuprates can be addressed in a single sample.

Here, we demonstrate a combined study on nearly optimally doped BSCCO by transport and tunneling spectroscopy. We stamp exfoliated BSCCO flakes on pre-patterned electrodes and observe superconductivity with a transition of 87 K in unprotected flakes as thin as four unit cells (4 UC). We further identify that the 4 UC flake contains only 1 UC that is superconducting. We achieve a reversible SIT in BSCCO flakes on SIC. Such a significant carrier modulation is realized by the electric field driven lithium ion intercalation. Scaling analysis shows that the SIT data in the quantum critical region collapse onto a single function, in agreement with a 2D-QPT. Tunneling spectroscopy in the G/BSCCO heterojunctions further unveils the evolution of the density of states across the SIT. Our technique constitutes an *in-situ* electron-doping method for addressing the complex phase diagram of cuprates.

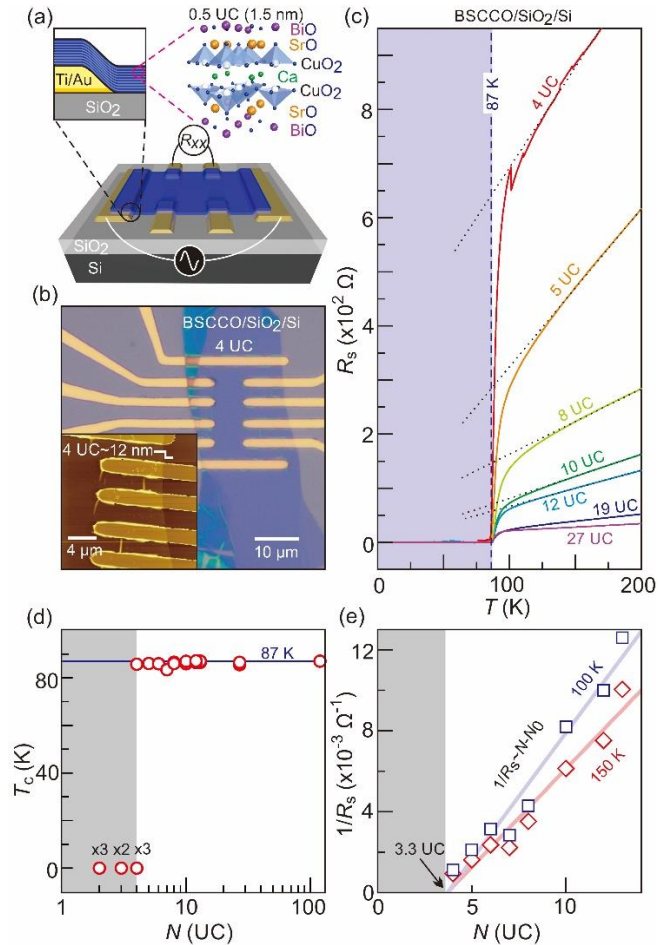

**Figure 1.** (a) Schematic drawing of the sample structure. The BSCCO flake is transferred on top of the Ti/Au (5/30 nm) electrodes. Upper-right illustrates the BSCCO lattice in half

of a unit cell. (b) Optical image of a 4 UC BSCCO. Inset shows the atomic force microscopy (AFM) image taken on the same sample. The film thickness from the step height is determined to be 12 nm (indicated by the solid line). (c) Sheet resistance of BSCCO flakes with different thicknesses. Dotted straight lines are guide to the eye. (d) Superconducting transition temperature as a function of sample thickness (number of UC). Numbers in the lower-left of the panel indicate multiple samples. (e) Thickness dependence of the inverse sheet resistances at 100 and 150 K. Straight lines are linear fits.

Retaining superconductivity in exfoliated BSCCO thin flakes is highly nontrivial[18–20]. So far, superconductivity in exfoliated BSCCO flakes was only achieved by covering them with graphene[20]. A possible factor for the degradation into an insulator is the chemicals used in sample processing. To avoid that, we stamp BSCCO flakes directly onto pre-patterned electrodes via a completely dry transfer method[21]. Figure 1(a) shows schematically the sample structure. Figure 1(b) provides the optical and atomic force microscopic images of a 4 UC flake on a $SiO_2$/Si substrate (Optical images of other samples are shown in the supporting information Figure s1). The temperature dependent resistivity of this 4 UC sample and that of thicker flakes ($N$ =5, 8, 10, etc.) are gathered in Figure 1(c). These flakes show the characteristic superconducting transition of BSCCO: the sheet resistance ($R_s$) plummets at 87 K from a linear trace (dotted lines in the inset of Figure 1(c)) at high temperatures. By simply avoiding the wet lithography, one can achieve superconducting flakes without any protection layers.

As shown in Figure 1(d), we obtain nearly the same $T_c$ for BSCCO on $SiO_2$/Si substrates, if $N \geq 4$. Four UC seems to be the minimum thickness for observing superconductivity in our experiment. To understand this sudden suppression, we analyze the inverse sheet resistance as a function of $N$. As shown in Figure 1(e), we obtain a linear dependence of $1/R_s$ on $N$. This trend is expected if the total conduction sums up identical contributions from individual layers. Interestingly, there exists an intercept $N_0 = 3.3$ such that: $1/R_s \propto N - N_0$. The exfoliated flakes therefore share a common number of dead layers, which is likely caused by the air exposure during sample transfer (see Supporting Information Figure s2). Samples with $N<N_0$ do not host superconductive layers, giving rise to the step like behavior of $T_c$ on $N$ (Figure 1(d)). Because of the dead layers, the superconducting 4

UC sample in Figure 1 (c) in fact possesses only 1 UC that hosts zero resistance at low temperatures.

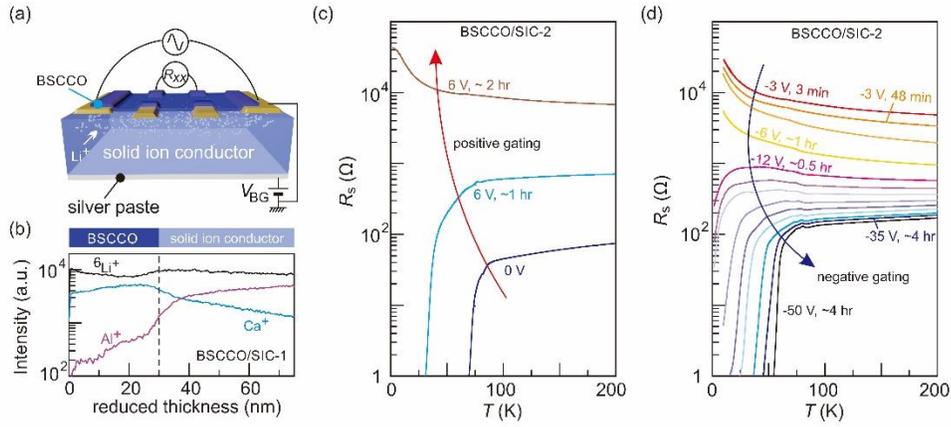

**Figure 2.** (a) Schematic drawing of the sample and the gating induced lithium intercalation process. (b) Depth dependence of the element selective intensities obtained by TOF-SIMS. This sample was gated at 6 V for 10 hours before doing TOF-SIMS. We integrate the signal from an area of about 1000 μm² in the central part of the BSCCO flake. The signal of $Ca^+$ ($Al^+$) originates from the BSCCO (SIC) only. (c)(d) Sheet resistances as a function of temperature for a 40 nm thick BSCCO flake tuned across the SIT. The leakage current at 300 K is below 5 nA at -50 V.

We now employ the solid ion conductor–a lithium containing ceramic– as the substrate[15, 22]. Figure 2(a) illustrates the device structure and the working principle. By ramping up back gate voltage ($V_{BG}$) to about 6 V, lithium ions are driven into BSCCO. This intercalation results in dramatic increase of the sample resistance, as will be discussed later. To verify the presence of lithium in BSCCO after intercalation, we employ the time-of-flight secondary ion mass spectroscopy (TOF-SIMS). This technique mills the sample in the direction perpendicular to substrate surface and analyzes the ions bombarded out of the material. Figure 2(b) plots the distribution of $^6Li^+$, $Ca^+$ and $Al^+$ in sample BSCCO/SIC-1 where the BSCCO flake is gated at 6 V for 10 hours. The drop of $Ca^+$ and the rise of $Al^+$ intensity indicate the boundary between BSCCO and SIC (dashed line). The almost constant signal of $^6Li^+$ across the boundary suggests that the density of lithium ions in BSCCO reaches the same level as that in the substrate.

Figure 2(c)(d) show the transport results of a 40 nm thick BSCCO (13 UC) under the solid state gating. The as-cleaved sample possesses a $T_c$ of 70 K (Figure 2(c)). The reduction of $T_0$ in comparison to samples shown in Figure 1 stems from the spontaneous diffusion of lithium from the solid ion conductor (see Supporting Information Figure s3). We apply back gate voltages at 300 K and wait for certain periods of time by monitoring the change of the resistance. The field induced movement of ions in the substrate can be terminated by cooling the sample down below 240 K (as indicated by the suppression of the gate leakage current). By applying $V_{BG} = 6$ V with consecutively increasing amount of time, the BSCCO flake transforms to the underdoped and finally to the insulating regime (Figure 2(c)). Reversing the applied gate voltages can recover the sample back to a superconductor, as shown by curves in Figure 2(d) (Details for the gating are given in the supporting information Figure s4). The superconducting temperature can be tuned back to about 50 K.

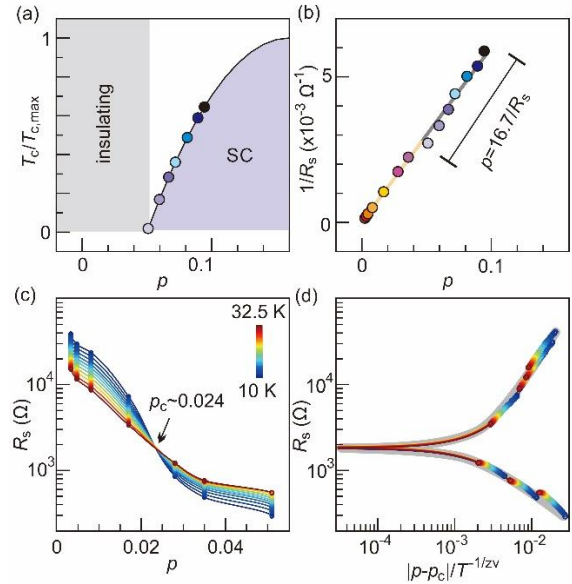

**Figure 3.** (a) Superconducting temperature $T_c$ normalized by the maximal $T_{c,max}$ plotted against the estimated doping level $p$. (b) Inverse sheet resistance ($1/\underline{R}_s$) as a function of the estimated doping level across the SIT. The bar marks the data points in the superconducting

regime which are used for the extrapolation into the insulating regime. (c) $R_s$ close to the SIT as a function of doping. (d) $R_s$ around the SIT as a function of $|p - p_c|T^{-1/zv}$.

In order to analyze the SIT quantitatively, we first estimate the corresponding doping level $p$ for each curves in Figure 2(d). Data in the superconducting regime are used to determine $p$ from $T_c$ by using the empirical relation: $T_c = T_{c,max} [1 - 82.6(p - 0.16)^2]$ [23] (Figure 3(a)). We then obtain the relation between $1/R_s(T=150$ K$)$ and $p$ in the superconducting regime: $p = 16.7 / R_s(T=150$ K$)$ [5, 6]. Such a linear dependence is then extrapolated to the insulating regime for the evaluation of $p$ there, as summarized in Figure 3 (b). Figure 3(c) replots the temperature dependent sheet resistance data in the quantum critical region of the SIT as a function of $p$. The critical doping $p_c$ is determined to be 0.024, above which the sample shows superconducting transition. According to the theory of 2D-QPT[5], data around the SIT is governed by a single function of $u = |p - p_c|T^{-1/zv}$. Figure 3(d) displays the resistance data just in this fashion, where the exponent $zv$ is fitted to be 1.5 ±0.2. The critical resistance here is about 2 kΩ. It is smaller than $h/4e^2$–expected from the QPT of a purely bosonic model. Similar observation has been seen in electron doped cuprates[9]. It was suggested that the SIT may involve fermionic excitations—apart from the phase fluctuations of Cooper pairs. For some samples, the temperature dependent resistance after lithium intercalation show kinks in the superconducting transition (Figure 4 (c)), indicating phase separation. Therefore, disorder in our system may generate intermediate states in the SIT [24], causing the deviation from an ideal 2D-QPT.

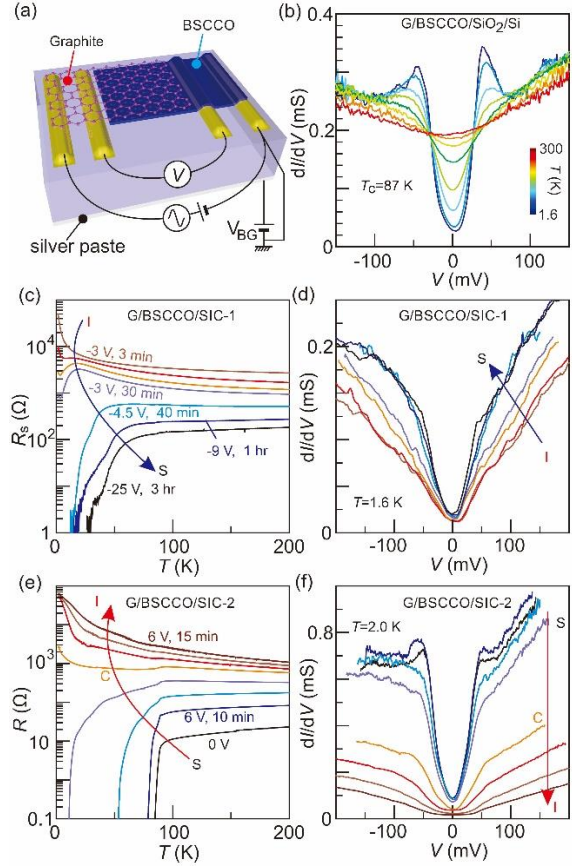

**Figure 4.** (a) Schematic drawing of the G/BSCCO heterostructure with the measurement configuration for the tunneling spectroscopy. (b) Tunneling conductance obtained from G/BSCCO/SiO$_2$/Si at different temperatures ($T$=1.6, 40, 70, 85, 110, 150, 200 and 300 K). This sample (50 nm) is realized on a SiO$_2$/Si, such that $T_c$=87 K. (c)(e) Resistances of BSCCO in G/BSCCO/SIC-1 and G/BSCCO/SIC-2 across the SIT. Data in panel (c) are sheet resistances. (d)(f) Tunneling spectra obtained for the two heterojunctions across the SIT of BSCCO. Traces in (c)(e) and (d)(f) with the same colors represent the same gated states. The thicknesses of the BSCCO flakes are 40 nm for (d)(e) and 100 nm for (e)(f), respectively. The four upper/lower traces in (e)/(f) are from a second pair of contacts

We further carry out spectroscopic studies by placing a graphite flake on top of BSCCO to form a tunnel junction[20, 25] (Figure 4(a)). Figure 4(b) displays the tunneling conductance across the first G/BSCCO heterojunction (G/BSCCO/SiO$_2$/Si). It reveals essentially the gap structure of BSCCO. For this sample with $T_c$ = 87 K, we obtain a gap of about 41 meV.

This gap fully vanishes at around 200 K, consistent with the pseudo-gap behavior for bulk BSCCO[26]. After verifying this probing scheme, we combine gating and tunneling on the other two devices fabricated on SIC. Figure. 4(c)-(f) show the SIT in transport of G/BSCCO/SIC-1, G/BSCCO/SIC-2 and the corresponding tunneling spectra, respectively.

This combination allows us to address the density of states continuously from the superconducting state to the insulating regime. Here we focus on the critical and insulating regimes, in which spectroscopic studies remain challenging and scarce[10, 27]. Figure 4(c) indicates that the resistance curve evolves from downward bending to shooting upward with a critical sheet resistance of about 5 k$\Omega$. In this critical regime, the density of states acquires a symmetric V-shaped gap (Figure 4(d)). Such a gap feature has been seen by scanning tunneling microscopy (STM) on insulating $Ca_2CuO_2Cl_2$[10]. In the STM study on $Ca_2CuO_2Cl_2$, the V-shaped gap was observed to be closely correlated with the charge order. This behavior is again seen in the thicker BSCCO sample—G/BSCCO/SIC-2 (marked as curve C in Figure 4 (e)(f)). It suggests that this V-shaped gap in the critical regime may be a universal feature for cuprates. In Figure 4(e)(f), we probe further into the insulating regime. There, the density of states on both sides of the Fermi level gets strongly suppressed. Interestingly, d$I$/d$V$ starts to become flat around zero bias. It seems to indicate that the pseudo-gap gets replaced by an insulating gap.

To summarize, we successfully exfoliate, without protection, superconducting BSCCO flakes down to 4 UC. By using the solid ion conductor, we intercalate lithium into BSCCO flakes and induce a reversible SIT. The evolution of the density of states across the SIT is probed by a G/BSCCO heterojunction. This work demonstrates that the solid ion conductor can be applied to 2D materials and it allows tunneling spectroscopy to be carried out on the same sample across the various quantum phases.

**Methods**

Bulk crystals of nearly optimally doped BSCCO ($T_c$=87 K) were grown by the traveling floating zone method[28]. They were mechanically exfoliated in the glovebox with Ar atmosphere ($H_2O$<0.1 ppm, $O_2$ <0.1 ppm). Dry transfer technique[21] was employed to deposit the flake onto prepatterned electrodes (Ti/Au: 5/30 nm). All the samples discussed in the main text were wire-bonded and then loaded into the cryogenic system with less than one hour of air exposure. To measure the resistances, we used the standard lock-in technique with a typical AC current of 1 μA at 13.33 Hz. The tunneling spectra were obtained by sending a DC+AC signal through the G/BSCCO heterojunction and measuring the DC and AC voltages in a four-terminal configuration. Back-gating was realized by applying DC gate voltages to the silver paste that was used to glue the solid ion conductor onto the chip carrier.

ASSOCIATED CONTENT

**Supporting Information**. This material is available free of charge via the Internet at http://pubs.acs.org. The optical images of BSCCO films with different thickness, *I-V* characteristics of G/BSCCO/SiO$_2$/Si junctions, extended data on the solid ion conductor gating of BSCCO, and details of the gating process.

AUTHOR INFORMATION

**Corresponding Author**

# dingzhang@mail.tsinghua.edu.cn

# qkxue@mail.tsinghua.edu.cn

**Author Contributions**

The manuscript was written through contributions of all authors. All authors have given approval to the final version of the manuscript.

**Notes**

The authors declare no competing financial interests.

ACKNOWLEDGMENT

We thank Yayu Wang, Matteo Minola, Bernhard Keimer for fruitful discussions. We are grateful to Lu Yang and Zhanping Li for technical assistance. This work is financially

supported by the National Natural Science Foundation of China (grant No. 11634007, 11790311, 11427903, 11604176); the Ministry of Science and Technology of China (2017YFA0304600, 2017YFA0302902, 2015CB921001); and the Beijing Advanced Innovation Center for Future Chip (ICFC). Work at Brookhaven is supported by the Office of Basic Energy Sciences, Division of Materials Sciences and Engineering, U.S. Department of Energy under Contract No. DE-SC0012704.

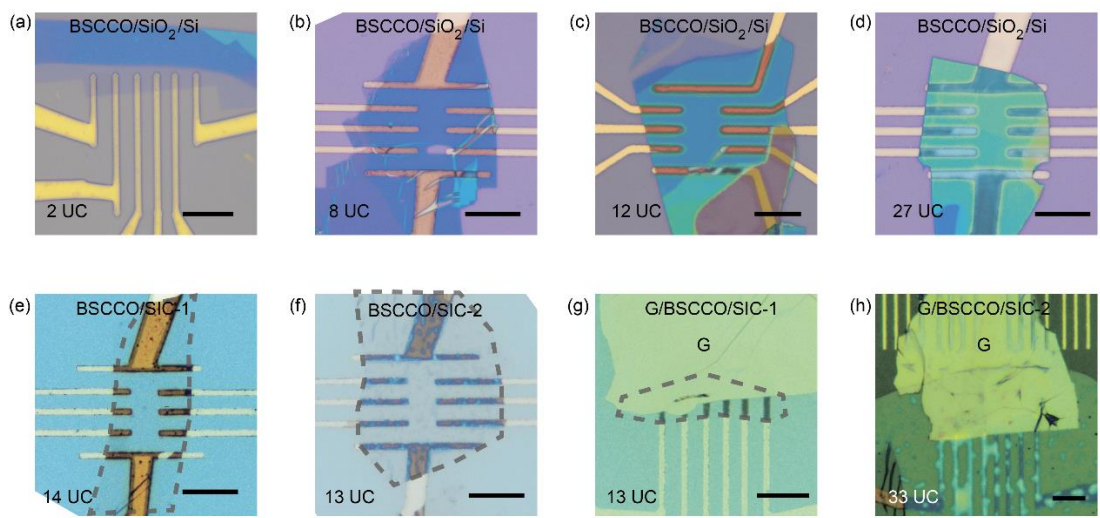

Figure s1 Optical images of BSCCO samples. Scale bars are all 15 μm. (a)-(d) BSCCO flakes on SiO$_2$/Si substrates. (e)(f) BSCCO flakes on the solid ion conductor. (g)(h) G/BSCCO heterojunctions on SIC. Dashed lines in (e)-(g) demarcate the boundaries of the BSCCO flakes.

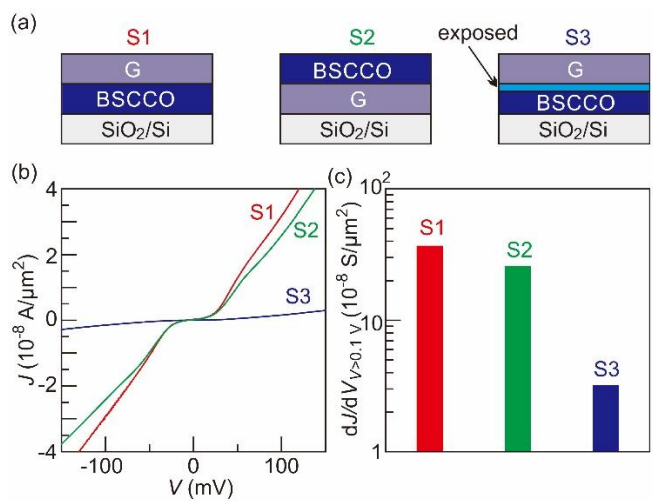

Figure s2 (a) Schematic illustrations of three G/BSCCO junctions: (1) G on BSCCO; (2) G beneath BSCCO; (3) G on exposed BSCCO. Samples 1 and 2 were fabricated in the glove box with Ar atmosphere. The junction areas are 720 and 600 μm$^2$, respectively. For sample 3, we exfoliated BSCCO in the glove box first and exposed the sample in air for about 1 hour. We then reloaded it into the glove box and transferred G on its top. (b) Comparison of the transport across the three heterojunctions. Sample 1 and 2 show the same behavior while sample 3 exhibits much higher resistance (lower conductance) across the junction. (c) Conductances of the three junctions at large bias. It indicates that a thick insulating layer forms when the BSCCO is exposed to air.

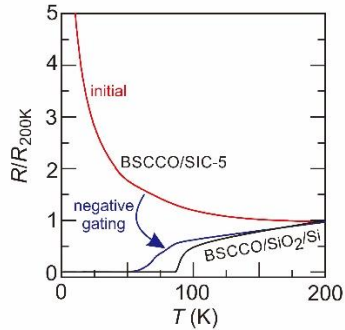

Figure s3 Normalized resistivity data of an as-transferred BSCCO flake on the solid ion conductor in comparison to that of the BSCCO flake exfoliated on the $SiO_2$/Si substrates. This sample on the solid ion conductor shows insulating behavior in the beginning even without applying any gate voltages. After negative gating, it resumes to be superconducting. This process indicates that lithium can spontaneously diffuse into BSCCO to suppress superconductivity.

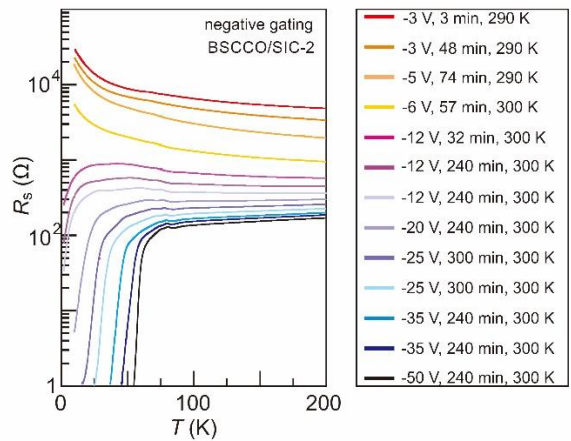

Figure s4 Sheet resistance data across the SIT with the detailed gating processes noted in the legend.

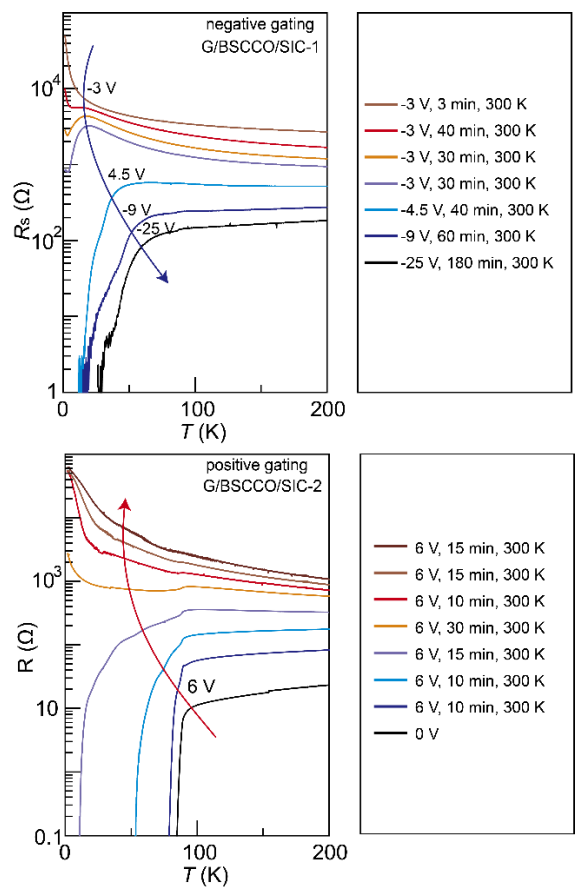

Figure s5 Sheet resistance data across the SIT with the detailed gating processes noted in the legend.